\documentclass[letterpaper]{article} 
\usepackage{aaai23}  
\usepackage{times}  
\usepackage{helvet}  
\usepackage{courier}  
\usepackage[hyphens]{url}  
\usepackage{natbib}  
\usepackage{algorithmic}

\usepackage{newfloat}
\usepackage{listings}

\usepackage[utf8]{inputenc} 
\usepackage[T1]{fontenc}    
\usepackage{url}            
\usepackage{booktabs}       
\usepackage{amsfonts}       
\usepackage{nicefrac}       
\usepackage{microtype}      
\usepackage[page,title]{appendix}
\usepackage{comment}
\usepackage{enumitem}
\usepackage{threeparttable}

\usepackage{xcolor}
\usepackage{bm}
\usepackage{dsfont}
\usepackage{amsmath}
\usepackage{amssymb}
\usepackage{amsthm}
\newtheorem{theorem}{Theorem}

\usepackage{graphicx}
\usepackage{float}
\usepackage{subfigure}
\usepackage{capt-of}
\usepackage{caption}

\usepackage{comment}

\usepackage{wrapfig}       
\usepackage{diagbox}
\usepackage{multirow}
\usepackage{makecell}

\usepackage{pifont}
\usepackage[ruled]{algorithm2e}

\def \cE {\mathcal{E}}
 
\def \P{\mathbb{P}} 

\def \bfE {\mathbb{E}}

 \def \cI {\mathcal{I}}
\def \cL{\mathcal{L}}
\def \cO{\mathcal{O}}
 \def \cD{\mathcal{D}}

\DeclareCaptionStyle{ruled}{labelfont=normalfont,labelsep=colon,strut=off} 
\floatstyle{ruled}
\newfloat{listing}{tb}{lst}{}
\floatname{listing}{Listing}
\title{Multiple Robust Learning for Recommendation}
\author{
Haoxuan Li\textsuperscript{\rm 1}\equalcontrib, Quanyu Dai\textsuperscript{\rm 2}\equalcontrib, Yuru Li\textsuperscript{\rm 2}, Yan Lyu\textsuperscript{\rm 1}, Zhenhua Dong\textsuperscript{\rm 2}, Xiao-Hua Zhou\textsuperscript{\rm 1}, Peng Wu\textsuperscript{\rm 3,1}\thanks{Peng Wu is the corresponding author of this paper.}}

\affiliations{
    \textsuperscript{\rm 1}Peking University\\
 \textsuperscript{\rm 2}Huawei Noah's Ark Lab\\
 \textsuperscript{\rm 3}Beijing Technology and Business University\\
 \{hxli, lyuyan\}@stu.pku.edu.cn, \{daiquanyu, liyuru3, dongzhenhua\}@huawei.com, \\ azhou@bicmr.pku.edu.cn, pengwu@btbu.edu.cn
}

\usepackage{bibentry}

\begin{document}

\maketitle

\begin{abstract}
 In recommender systems, a common problem is the presence of various biases in the collected data, which deteriorates the generalization ability of the recommendation models and leads to inaccurate predictions.
 Doubly robust (DR) learning has been studied in many tasks in RS, with the advantage that unbiased learning can be achieved when either a single imputation or a single propensity model is accurate. In this paper, we propose a multiple robust (MR) estimator that can take the advantage of multiple candidate imputation and propensity models to achieve unbiasedness. Specifically, the MR estimator is unbiased when any of the imputation or propensity models, or a linear combination of these models is accurate. Theoretical analysis shows that the proposed MR is an enhanced version of DR when only having a single imputation and propensity model, and has a smaller bias. Inspired by the generalization error bound of MR, we further propose a novel multiple robust learning approach with stabilization. We conduct extensive experiments on real-world and semi-synthetic datasets, which demonstrates the superiority of the proposed approach over state-of-the-art methods.
\end{abstract}

\section{Introduction}
Recommender systems (RS) have made remarkable progress on many tasks in recent years, such as rating prediction with explicit feedback~\cite{IPS,DRJL}, user-item interaction prediction with implicit feedback~\cite{yang2018unbiased,saito2019unbiased,crmf}, post-view click-through rate prediction~\cite{CIKM-YuanHYZCDL19}, post-click conversion rate prediction~\cite{Zhang-etal2020,MRDR}, and uplift modeling~\cite{Sato-Singh2019,Sato-Takemori2020}.   Among them, an important challenge is that the collected data {are} always not the representative  
 of the overall population of interest, due to the presence of selection bias and confounding bias~\cite{Chen-etal2020, Wu-etal2022}. To address this problem, recent debias-related studies design causality-inspired approaches to achieve unbiased learning for RS. 
 Specifically,~\citet{marlin2012collaborative} and \citet{steck2013evaluation} proposed error imputation-based (EIB) approaches, which leverage an additional imputation model to estimate the prediction errors for missing data. However, the imputed errors tend to be inaccurate in practice and will probably cause a large bias. 
 Besides, \citet{IPS} proposed the inverse propensity score (IPS) and self-normalized inverse propensity score (SNIPS) methods for unbiased learning with the assistance of a propensity model. 
 Furthermore, \citet{DRJL} proposed a doubly robust (DR) estimator to combine an error imputation model and a propensity model. DR has double robustness, which is unbiased if either the imputed errors or propensities are accurate.
 Currently, DR and its enhanced versions~\cite{MRDR, Dai-etal2022, ding2022addressing} achieve the state-of-the-art performance for debiasing recommendation tasks.
 
The strength of doubly robust estimators is attributed to the two chances of achieving unbiased learning, i.e., accurate estimation of error imputation model or propensity model. 
However, the correct specifications of these models are demanding. If both of the learned models are mildly misspecified, the bias of the doubly robust learning can be severe as pointed out by existing work~\cite{Tan2007, Wu-Han2022}.   
What is worse, it is challenging to accurately learn them based on observed data, even if there are correct model specifications.
Firstly, the accurate imputed errors are hard to obtain, since the imputation model is learned with the exposed events while being used for the unexposed events. This can easily cause a large bias if the distributions of the exposed events and unexposed events are significantly different~\cite{Dai-etal2022}.
Secondly, the collected data always contains many biases~\cite{Chen-etal2020,Wu-etal2022} (e.g., biases induced by unobserved confounders), which may result in the inaccurate estimation of propensities.
Therefore, more research efforts for unbiased  learning are needed.

In this paper, we propose a multiple robust (MR) estimator for unbiased learning in RS, which contains multiple candidate propensity models and error imputation models. It allows multiple different specifications and learning of the propensity and imputation models.
Through theoretical analysis, we demonstrate that the proposed MR estimator has multiple robustness, which is unbiased if any of the propensity models, imputation models, or even a linear combination of these models can accurately estimate the true propensities or prediction errors. 
Therefore, the MR estimator can significantly tackle the problem of inaccuracy in learned propensities or imputed errors from a single model of existing work, by providing more opportunities for model specification and learning.
In addition, we particularly analyze the relation between MR and DR when MR has only one propensity model and one imputation model, and find that our MR estimator can actually provide an enhanced version of double robustness.
By analyzing the tail bound of the MR estimator, we further propose a multiple robust learning method with stabilization, to better control the generalization error by {adding} $L_2$ regularization. 
Extensive experiments on both real-world and semi-synthetic datasets demonstrate that the proposed MR method achieves a significant improvement compared to state-of-the-art methods. 

\section{Related Work}  
{\bf Unbiased Learning in Recommendation.}  Due to the inconsistency between the collected data and the target population, RS is always affected by various biases~\cite{Wu-etal2022}. Therefore, unbiased learning has been extensively studied~\cite{marlin2009collaborative, Steck2010, DIB}. 
In RS, \citet{IPS} and \citet{saito2019unbiased} used a single propensity model for unbiased learning on explicit and implicit feedback tasks,  
while~\citet{marlin2012collaborative} and  \citet{steck2013evaluation} adopted a single error imputation model for  unbiased learning. 
\citet{DRJL} proposed to combine a propensity and an error imputation model to achieve 
doubly robust learning, and yielded superior performance on ranking~\cite{saito2020doubly, kiyohara2022doubly}, uplift modeling~\cite{Sato-Takemori2020} and data fusion~\cite{bonner2018causal,liu2020general,Chen-etal2021,Wang-etal2021}.  
Based on that, \citet{MRDR}, \citet{Dai-etal2022}, \citet{TMLE} and \citet{SDR} proposed more robust DR methods, and \citet{Zhang-etal2020} proposed a parameter sharing mechanism applied to multi-task learning. 
In addition, \citet{ding2022addressing} proposed a robust deconfounder framework to mitigate  the influence of unmeasured confounder. 
Different from the existing debiasing methods, this paper develops a novel multiple robust learning approach, which allows multiple propensity and error imputation models and thus enables a greater chance of unbiased learning. 

\noindent 
{\bf Multiple Robust.} 
MR approaches were designed to prevent the inaccuracy of learned propensities and imputed errors.
Currently, the MR methods mainly focused on estimating the population mean of a response variable with missing~\cite{Li-etal2020}. They can be  divided into two major categories: empirical likelihood-based calibration  methods~\cite{Han-Wang2013, Chen-2017, Han-2016, Han-2018} and
  ensemble learning-based methods~\cite{Chan-2013, Chan-2014, Duan-Yin2017}.    
The former relies on solving a large optimization problem and is hard to implement in the data of RS.   To ease the computational burden, the latter decomposes the optimization problem into two steps,  
which first compress multiple propensity/imputation models into one new propensity/imputation model, and then solve an optimization problem the same as the calibration approaches~\cite{Duan-Yin2017}. How to reduce the computational burden and how to embed them in a joint learning model in RS are challenges problems.   
    To the best of our knowledge, this is the first work that proposes a multiple robust learning approach for debiased recommendations.

\section{The Proposed Method} 

\subsection{Problem Definition and Preliminary} 
Let $\mathcal{U} =\{u\}$ be the users, $ \mathcal{I} = \{i\}$ the items, and $\mathcal{D} = \mathcal{U} \times \mathcal{I}$ all of the user-item pairs.
Let $x_{u,i}$ be the feature of a user-item pair $(u,i)$, which is used by the prediction model $f(x_{u,i}; \theta)$ with model parameters $\theta$ in RS, to predict the item $i$ rating $y_{u,i}$ for user $u$. 
Denote $y_{u,i}$ as the true rating of a user $u$ on an recommended item $i$. In RS, a large fraction of $y_{u,i}$ is missing-not-at-random, resulting in the highly sparse and biased data along with the challenge of unbiased model learning.
Define $o_{u,i}$ as an indicator function of whether $y_{u,i}$ is observed or not, i.e., $o_{u,i}=1$ if $y_{u,i}$ is observed and $o_{u,i}=0$ otherwise. Denote $\mathcal{O}=\{(u,i)\in \mathcal{D}: o_{u,i}=1\}$ as the indices of observed samples, and $\hat{\mathbf{Y}}$ as the predicted score matrix of all user-item pairs, obtained from prediction model $f(x_{u,i}; \theta)$. This work aims to train the prediction model based on the observed samples to infer all true ratings $y_{u,i}$.

%
To evaluate the performance of the prediction model $f(x_{u,i}; \theta)$ over all user-item pairs, let $e_{u,i}$ be the prediction error, which is an appropriately chosen loss function, e.g., the squared loss $(y_{u,i} - f(x_{u,i}; \theta))^2$. However, the unbiased evaluation on $\mathcal{D}$ requires all values of $y_{u,i}$. 
Given all the predicted ratings $\hat y_{u,i}\triangleq f(x_{u,i}; \theta)$, ideally, if all ratings $y_{u,i}$ are observed, 
the prediction model $f(\cdot; \theta)$ can be obtained by directly optimizing the following ideal loss
    \[   \mathcal{E}_{ideal} =   \mathcal{E}_{ideal}(\hat{\mathbf{Y}}) = |\cD|^{-1} \sum_{(u,i)\in \cD} e_{u,i}.    \] 
However, since $y_{u,i}$ is observed only when $o_{u,i} = 1$, the ideal loss is non-computable. To address this problem, many debiasing methods in RS have been proposed by constructing unbiased substitutes of the ideal loss, such as the EIB estimator $\mathcal{E}_{EIB} = |\cD|^{-1} \sum_{(u,i)\in \cD} [ o_{u,i} e_{u,i} + (1-o_{u,i}) \hat m_{u,i}]$ and IPS estimator $\mathcal{E}_{IPS} = |\cD|^{-1} \sum_{(u,i)\in \cD} o_{u,i} e_{u,i} / \hat p_{u,i}$, 
where $\hat m_{u,i} \triangleq m(x_{u,i}; \hat \beta)$ is an estimate of the prediction error $e_{u,i}$ from an error imputation model with parameter $\hat \beta$, and $\hat p_{u,i} \triangleq \pi(x_{u,i}; \hat \alpha)$ is an estimate of propensity score $p_{u,i} \triangleq \P(o_{u,i}=1 | x_{u,i})$ from a propensity model with parameter $\hat \alpha$. 
Besides, the doubly robust (DR) uses both imputed errors and propensities to further reduce the bias, which is 
    \[
        \mathcal{E}_{DR} = |\cD|^{-1} \sum_{(u,i) \in \cD} \Big [ \hat m_{u,i}  +  \frac{ o_{u,i} (e_{u,i} -  \hat m_{u,i} ) }{ \hat p_{u,i} } \Big ]. 
    \]

\subsection{Multiple Robust Estimator}
The unbiasedness of IPS and EIB estimators rely on the accurate estimations of propensities and prediction errors, respectively, from a single model. 
The DR estimator takes advantage of both the estimated propensities and imputed errors to achieve double robustness, but it only allows to use one propensity model and one imputation model. 
Nevertheless, the challenges of model specification and model learning significantly hinder the realization of unbiased learning. 
In this work, we propose a multiple robust (MR) estimator to tackle these challenges by introducing multiple propensity models and error imputation models. 
 Specifically, consider $J$ propensity models and $K$ imputation models 
\begin{align*}
            \mathcal{G}={}& \{ \pi_1(x; \hat \alpha_{1}), \dots, \pi_J(x; \hat \alpha_{J})\}, \\
            \mathcal{M}={}& \{ m_{1}(x; \hat \beta_{1}), \dots, m_{K}(x; \hat \beta_{K})\}. 
        \end{align*}
 Let $\hat p_{u,i}^j \triangleq \pi_j(x_{u,i}; \hat \alpha_{j})$ and $\hat m_{u,i}^k \triangleq m_{k}(x; \hat \beta_{k})$. 
    \[  \boldsymbol{u}(x_{u,i}) = \left( 1/ \hat p^{1}_{u,i}, \cdots, 1/\hat p_{u,i}^J, \hat m_{u,i}^1 , \cdots, \hat m_{u,i}^{K} \right)^{T}.     \]  
The proposed MR estimator is given as 
        \begin{equation}  \cE_{MR}  =  |\cD |^{-1} \sum_{(u,i) \in \cD}  \boldsymbol{u}^{T}(x_{u,i}) \cdot \hat{\boldsymbol{\eta}}(\theta),
                   \end{equation} 
where $\hat{\boldsymbol{\eta}}(\theta)$ is the solution by minimizing
\begin{equation} \label{eq-ls} \sum_{(u,i)\in \mathcal{D}} o_{u,i} \{e_{u,i}-  \boldsymbol{u}^{T}(x_{u,i}) \cdot \boldsymbol{\eta} \}^2.\end{equation}    
Denote $ \hat{\boldsymbol{\eta}}(\theta)$ as $\hat{\boldsymbol{\eta}}$ for simplification. Let $[\hat{\mathbf{P}}^j]_{u,i}=1/\hat p_{u,i}^j$ be the $(u,i)$-th element of the $j$-th learned inverse propensities matrix, and $[\hat{\mathbf{E}}^k]_{u,i}=\hat m_{u,i}^k$ be the $(u,i)$-th element of the $k$-th imputed errors matrix.  
\begin{theorem}[Multiple Robustness] \label{th1} 
MR is consistent\footnote{An estimator is consistent if, as the sample size increases, it converges to the true value of the parameter in probability.} 
when either of the following conditions hold: 
  
  (a) there exists a linear combination of the $J$ inverse propensities accurate, i.e., $ [\hat{\mathbf{P}}^{ln}]_{u,i}=1/p_{u,i}$;
  
  (b) there exists a linear combination of the $K$ imputed errors accurate, i.e. $ [\hat{\mathbf{E}}^{ln}]_{u,i}=e_{u,i}$,\\  
where $\hat{\mathbf{P}}^{ln}=\sum_{j=1}^{J}w_{j}\hat{\mathbf{P}}^j$ and $\hat{\mathbf{E}}^{ln}=\sum_{k=1}^{K}v_{k}\hat{\mathbf{E}}^k$ are the linear combinations of $\hat{\mathbf{P}}^j$ and $\hat{\mathbf{E}}^k$.

 In addition, the MR estimator $\cE_{MR}$ is unbiased, if $\hat{\boldsymbol{\eta}}$ and $\cE_{MR}$ are obtained through different samples. 
\end{theorem} 


We formally elaborate the multiple robustness of the MR estimator. Firstly, note that $\hat{\boldsymbol{\eta}}$ will make the first-derivative of equation (\ref{eq-ls})  with respect to $\boldsymbol{\eta}$ 0, i.e.,  
        \[    \sum_{ (u,i)\in \cD } o_{u,i} \cdot [  e_{u,i} - \boldsymbol{u}^{T}(x_{u,i}) \cdot  \hat{\boldsymbol{\eta}} ] \cdot \boldsymbol{u}(x_{u,i})   = 0^{(J+K)\times 1},  \]
of which the $j$-th element is that 
       $\sum_{ (u,i)\in \cD }  \{  e_{u,i} - \boldsymbol{u}^{T}(x_{u,i}) \cdot  \hat{\boldsymbol{\eta}} \} o_{u,i}/\hat p_{u,i}^j  = 0$.   
Therefore, for any given $j = 1,..., J$,  
    \begin{align}
 \cE_{MR}    
        ={}& \frac{1}{ |\cD | } \sum_{(u,i) \in \cD}  \boldsymbol{u}^{T}(x_{u,i}) \cdot \hat{\boldsymbol{\eta}} + \notag \\
        & \underbrace{\frac{1}{ |\cD | } \sum_{ (u,i)\in \cD }  \frac{o_{u,i}}{\hat p_{u,i}^j} \cdot \{  e_{u,i} - \boldsymbol{u}^{T}(x_{u,i}) \cdot  \hat{\boldsymbol{\eta}} \}}_{0} \notag \\
        ={}&  \frac{1}{ |\cD | }  \sum_{ (u,i)\in \cD }  \frac{o_{u,i}}{\hat p_{u,i}^j} \cdot  e_{u,i}  + \notag \\
        {}& \frac{1}{ |\cD | } \sum_{ (u,i)\in \cD } \left[ 1 -  \frac{o_{u,i}}{\hat p_{u,i}^j}  \right] \cdot  \boldsymbol{u}^{T}(x_{u,i}) \cdot  \hat{\boldsymbol{\eta}}.\label{eq1}
    \end{align} 
Suppose $1/\hat p_{u,i}^j$ is an accurate estimate of $1/p_{u,i}$ and  it can be shown that the last term in equation (\ref{eq1}) will converge to 0 in probability, which implies that 
    \[ \bfE_{\mathcal{O}}[  \cE_{MR}] \stackrel{P}{\longrightarrow}  |\cD|^{-1} \sum_{(u,i)\in \cD}  \bfE_{\cO}[ o_{u,i} e_{u,i} /\hat p_{u,i}^j]    
    = \cE_{ideal}.        \]
If $[\hat{\mathbf{P}}^{ln}]_{u,i} = 1 / p_{u,i}$ for some weights $w_1,..., w_J$, the consistency of  $\cE_{MR}$ follows immediately from the weighted summation of the equation (\ref{eq1}) with different $j$.

Instead, if one of the imputed errors, 
 or a linear combination of the $K$ imputed errors can estimate $e_{u,i}$ accurately, i.e., $ [\hat{\mathbf{E}}^{ln}]_{u,i}=e_{u,i}$ for some weights $v_1, ..., v_K$,  
which implies that $\hat{\boldsymbol{\eta}} =  (0,\cdots, 0, v_1, ..., v_K)^T$,   
resulting in $\bfE_{\mathcal{O}}[  \cE_{MR}]  = \cE_{ideal}$.
Thus, $\cE_{MR}$ is multiple robust in the sense of consistency. In addition, 
when using different samples to calculate $\hat{\boldsymbol{\eta}}$ and $\cE_{MR}$, such independence ensures that the expectation of last term in equation (\ref{eq1}) is strictly equal to 0, resulting in $\bfE_{\mathcal{O}}[  \cE_{MR}]=\cE_{ideal}$ (See the next section for implementation details). 

Theorem \ref{th1} indicates that the MR estimator can achieve
unbiasedness under weaker conditions than those of the DR estimator. 
Instead of IPS and EIB relying only on a single model, and DR relying on a single propensity and a single imputation model, the MR estimator allows to exploit multiple candidate models to reach unbiasedness.
In addition, MR has great potential in industrial application scenarios of RS, since there are usually many candidate models for recommendation tasks, and MR further allows combining those models as shown in Theorem \ref{th1} and enhances the prediction accuracy. 

When the MR estimator contains exactly \emph{one error imputation model} $\hat m_{u,i}$ and \emph{one propensity model} $\hat p_{u,i}$, a natural question is the relationship between MR and DR. 
Theorem \ref{th2} states that MR acts as a boosted version of DR in such a case. 
\begin{theorem}[Relation to DR] \label{th2} Given one error imputation model and one propensity model, then

(a) (Enhanced double robustness) $\cE_{MR}$ has double robustness. Furthermore, when both the imputation model and propensity model are inaccurate, $\cE_{MR}$ retains unbiasedness in condition that $e_{u,i}$ can be linearly represented by $\hat m_{u,i}$ and $1/\hat p_{u,i}$, but $\cE_{DR}$ doesn't.

(b) (Equivalent Form) $\cE_{MR}=\cE_{DR}$ if the error imputation model is accurate. 
\end{theorem}  
 
Note that DR can be written as
{\small \begin{equation} \label{eq3}     \cE_{DR}   = \frac{1}{ |\cD | }  \sum_{ (u,i)\in \cD }   \biggl [  \frac{o_{u,i} \cdot   e_{u,i} }{\hat p_{u,i}} +   \Big \{ 1 -  \frac{o_{u,i}}{\hat p_{u,i}} \Big \} \cdot   \hat m_{u,i}    \biggr ],   \end{equation}}
and it follows from   
equation (\ref{eq1}) that  
{\small \begin{equation} \label{eq4}
 \cE_{MR}  
        =  \frac{1}{ |\cD | }  \sum_{ (u,i)\in \cD } \biggl [  \frac{o_{u,i}\cdot 
e_{u,i} }{\hat p_{u,i}}        +  \Big \{ 1 -  \frac{o_{u,i}}{\hat p_{u,i}} \Big \} \cdot  \boldsymbol{u}^{T}(x_{u,i}) \cdot  \hat{\boldsymbol{\eta}} \biggr ]. 
    \end{equation}}  
Therefore, $\cE_{MR}$ degenerates to $\cE_{DR}$ if $\hat m_{u,i}=\boldsymbol{u}^{T}(x_{u,i}) \cdot  \hat{\boldsymbol{\eta}}$, for which a sufficient condition is $e_{u,i} = \hat m_{u,i}$, i.e., the error imputation model is accurate.

By a comparison of equation (\ref{eq3}) and equation (\ref{eq4}), it can be seen that MR performs as a different estimator with double robustness, replacing the $\hat m_{u,i}$ with $\boldsymbol{u}^{T}(x_{u,i})\cdot \hat{\boldsymbol{\eta}}$.
Importantly, if $1/\hat p_{u,i}$ is helpful to fit $e_{u,i}$, then $\boldsymbol{u}^{T}(x_{u,i})\cdot \hat{\boldsymbol{\eta}}$ is expected to be a more accurate estimate  of the prediction error than $\hat m_{u,i}$. If $1/\hat p_{u,i}$ doesn't help,
 then $\boldsymbol{u}^{T}(x_{u,i})\cdot \hat{\boldsymbol{\eta}}$ and $\hat m_{u,i}$ will have the same accuracy.
 Thus, $\cE_{MR}$ can be viewed as an  improved version of DR when we only have access to an error imputation model and a propensity model.

Theorem \ref{th1} shows that the bias of MR vanishes with an accurate model. However, when all of the models are inaccurate, the bias of MR is given in Theorem \ref{th3}. 
\begin{theorem}[Bias of MR] \label{th3} 
Given the $J$ propensity models and $K$ imputation models,  
with $\hat p_{u,i}^j >0$ for all $(u,i)$ pairs, then the bias of MR estimator is given as   
\begin{align*}
    \operatorname{Bias}\left(\mathcal{E}_{MR}\right)
    ={}& \frac{1}{ |\cD | } \Biggl | \sum_{ (u,i)\in \cD }  \underbrace{ \Big\{ 1- p_{u,i} \sum_{j=1}^J \frac{w_j}{\hat p_{u,i}^j} \Big \}}_{\text {linear combination of}~1/\pi_1, \dots, 1/\pi_J} \times \\
    {}& \underbrace{ \left\{ e_{u,i} - \boldsymbol{u}^{T}(x_{u,i}) \cdot  \bfE_{\mathcal{O}}[\hat{\boldsymbol{\eta}}]\right\}}_{\text {linear combination of multiple models}} \Biggr | + O(|\cD|^{-1}),
\end{align*}
where $\sum_{j=1}^J w_j/\hat p_{u,i}^j$ is the best linear approximation of $1/p_{u,i}$.
\end{theorem}

Theorem \ref{th3} shows that the bias of MR estimator consists of a dominant term and a negligible term of order $O(|\cD|^{-1})$. 
In addition, it can be seen that the bias of MR has similar form compared to that of DR, based on the $j$-th propensity model and $k$-th imputation model, which is given by  
\begin{align*}
    \frac{1}{ |\cD | } \Big | \sum_{ (u,i)\in \cD }  \underbrace{\Big \{ 1-\frac{p_{u,i}}{ \hat p_{u,i}^j } \Big \}}_{\text {\emph{single propensity}}~1/\pi_j} \cdot \underbrace{\left\{e_{u,i}- \hat m_{u,i}^k\right\}}_{\text {\emph{single imputation}}~m_k} \Big |.
\end{align*} 
Note that both the biases consist of the product of the estimation inaccuracies from learned propensities and imputed errors. However, MR allows using the optimal linear combination of multiple propensity models to approximate $p_{u,i}$, as well as using the optimal linear combination of all models to approximate $e_{u,i}$, while DR only uses one single propensity and imputation model to approximate $p_{u,i}$ and $e_{u,i}$, respectively. This  shows that MR tends to have a smaller bias compared to DR.

\subsection{Alternating Multiple Robust Learning with Stabilization} \label{sec4.3}
In this subsection, we analyze the tail bound of the MR estimator and the generalization error bound of the learned prediction model based on the MR loss, and find that both of them depend on the norm of $\hat{\boldsymbol{\eta}}$, which motivates us to propose an alternative multiple robust learning with stabilization. The following Theorem \ref{th4} presents the tail bound of the MR estimator. 

\begin{theorem}[Tail Bound of MR] \label{th4} Suppose all $J$ propensity models satisfy $\Gamma^{-1} \leq \hat p_{u,i}^j \leq 1$ $(\Gamma \geq 1)$ for $j=1, \ldots, J$, and all $K$ imputation models satisfy $| \hat m_{u,i}^k | \leq M/2$~ for $k=1, \ldots, K$. Then for any prediction matrix $\hat{\mathbf{Y}}$ with given $\hat{\boldsymbol{\eta}}$, with probability $1-\delta$, the MR estimator does not deviate from its expectation by more than
{\small \[ \left|\cE_{MR}\left(\hat{\mathbf{Y}} \right)-{\bfE_{O}}\left[\cE_{MR}\left(\hat{\mathbf{Y}} \right)\right]\right| \leq \sqrt{\frac{\log ({2}/{\delta})}{2 { |\cD| }} } \max(\Gamma-1, M) \cdot \Vert \hat{\boldsymbol{\eta}} \Vert_1.
\]} 
\end{theorem}

For any given hypothesis space $\mathcal{H}$ of the prediction model, the optimal prediction model $\hat{\mathbf{Y}}^{\dagger}$ can be defined by
$
\hat{\mathbf{Y}}^{\dagger}= \arg \min_{\hat{\mathbf{Y}} \in \mathcal{H}} \mathcal{E}_{MR}(\hat{\mathbf{Y}}).
$
Then, the generalization error bound of $\hat{\mathbf{Y}}^{\dagger}$ from $\mathcal{H}$ obtained using MR can be further derived as shown in Theorem \ref{th5}. 
\begin{theorem}[Generalization Error Bound] \label{th5} For any finite hypothesis space of predictions $\mathcal{H}=\{\hat{\mathbf{Y}}_{1}, \ldots, \hat{\mathbf{Y}}_{|\mathcal{H}|}\}$, then under the conditions of Theorems \ref{th1} and \ref{th4}, the MR estimator deviates from the true risk $\mathcal{E}_{ideal}(\hat{\mathbf{Y}}^{\dagger})$ with given $\hat{\boldsymbol{\eta}}$ is bounded with probability $1-\delta$ by
{\small \[ 
\mathcal{E}_{ideal}(\hat{\mathbf{Y}}^{\dagger}) \leq  \cE_{MR}\left(\hat{\mathbf{Y}}^{\dagger} \right)+ \sqrt{\frac{\log (2|\mathcal{H}| / \delta)}{2 {|\cD|}}} \max(\Gamma-1, M) \cdot  \Vert \hat{\boldsymbol{\eta}} \Vert_1. 
\] }
\end{theorem} 
Theorem \ref{th5} shows that the generalization error bound can be controlled via the tighter range bounds of the propensity and imputation models, as well as a smaller value of $\Vert \hat{\boldsymbol{\eta}} \Vert_1$. The former can be achieved by clipping the extremely small estimated propensities or imposing an additional penalty for extreme imputed errors. The latter can be achieved by imposing an additional regularization term on the MR loss to penalize the norm of $\hat{\boldsymbol{\eta}}$ to stabilize the MR estimator.

A key design of the learning process is to use different samples to compute $\boldsymbol{\hat \eta}$ and update the prediction model with MR loss. As shown in Theorem \ref{th1}, such independence  
guarantees the strict unbiased learning process. Specifically, we first randomly sample a batch $\cD^\prime$ from $\cD$, and use the samples $\cO^\prime$ with observed ratings in $\cD^\prime$ to compute $\boldsymbol{\hat \eta}$. Next, another batch is drawn from the remaining samples $\mathcal{D}\setminus \cD^{\prime}$ for the calculation of MR loss and the update of the prediction model. 

Since Lasso has no closed-form solution 
and the iterative procedure will increase the computational cost, we take the Ridge regression with $\mathcal{O}^{\prime}\subset\mathcal{D}^{\prime}$ to obtain $\boldsymbol{\hat \eta}$ in MR
{\small \begin{equation}\label{eq:eta}
\boldsymbol{\hat \eta} = \Big [ \sum_{  (u,i)\in \mathcal{O}^{\prime}   } \boldsymbol{u}(x_{u,i})\cdot  \boldsymbol{u}^{T}(x_{u,i}) + \lambda \boldsymbol{I} \Big ]^{-1}  \Big [ \sum_{ (u,i)\in \mathcal{O}^{\prime} } \boldsymbol{u}(x_{u,i}) \cdot  e_{u,i}  \Big ],
\end{equation}}
where $\boldsymbol{I}$ is an identity matrix, and $\lambda$ is a hyper-parameter for stabilization.   
Then the prediction model is updated by the proposed MR loss $\cL_{MR}\left(\theta; {\hat \alpha}, {\hat \beta}\right)  =  \sum_{(u,i) \in \cD\setminus \cD^{\prime}}  \boldsymbol{u}(x_{u,i})^{T} \boldsymbol{\hat \eta}(\theta)$.  
In addition, given the prediction model and sampling data $\left\{\left(u_{k_l}, i_{k_l}\right)\right\}_{l=1}^{L}$ for $k=1,\dots,K$ from $\mathcal{O}$, the $k$-th imputation model is updated by using 
\[
\cL_{e_k}\left(\theta, {\beta_k}\right)=L^{-1} \sum_{l=1}^L ( m_k(x_{u_{k_l}, i_{k_l}};  \beta_k)-e_{u_{k_l}, i_{k_l}})^{2} / \hat p^{j}_{u,i},     
\] 
where $e_{u,i}=y_{u,i}-f_{\theta}\left(\boldsymbol{x}_{u,i}\right)$ and $p^{j}_{u,i}$ is a randomly-chosen propensity model. By alternatively updating the imputation models and prediction ratings with the above process, we can finally achieve unbiased learning with stabilization. We summarized the alternating learning approach in Alg. \ref{alg1}.
\begin{algorithm}[t]
\caption{Alternating Multiple Robust Learning with Stabilization} 
\label{alg1}
\LinesNumbered 
\KwIn{observed ratings $\mathbf{R}^{o}$, propensity models $\pi_1, \dots, \pi_j$, and stabilization parameter $\lambda$}
\While{stopping criteria is not satisfied}{
    \For{$k\in\{1, \dots, K\}$}{
    \For{number of steps for training the $k$-th imputation model}{Sample a batch of user-item pairs $\left\{\left(u_{k_l}, i_{k_l}\right)\right\}_{l=1}^{L}$ from $\mathcal{O}$\;
    Update $\beta_k$ by descending along the gradient $\nabla_{\beta_k} \cL_{e_k}\left(\theta, {\beta_k}\right)$
    }}
    \For{number of steps for training the prediction model}{Sample a batch of user-item pairs $\cD^{\prime}$ from $\cD$\;
    Obtain the rated samples in $\cD^{\prime}$ as $\left\{\left(u_{m}, i_{m}\right)\right\}_{m=1}^{M}= \mathcal{O}^{\prime}\subseteq \cO$\;
    $\eta\gets[ \sum_{  (u,i)\in \mathcal{O}^{\prime}   } \boldsymbol{u}(x_{u,i})\cdot  \boldsymbol{u}^{T}(x_{u,i}) +\lambda I]^{-1}  [ \sum_{ (u,i)\in \mathcal{O}^{\prime} } \boldsymbol{u}(x_{u,i}) \cdot  e_{u,i}  ]$\;
    Sample a batch of user-item pairs $\left\{\left(u_{n}, i_{n}\right)\right\}_{n=1}^{N}$ from $\mathcal{D}\setminus \cD^{\prime}$\;
    Update $\theta$ by descending along the gradient $\nabla_{\theta} \cL_{MR}\left(\theta; \alpha, \beta\right)$}
}
\end{algorithm}

\section{Experiments}
In this section, we conduct experiments on both real-world datasets and semi-synthetic datasets to evaluate the effectiveness of our proposed method.

\smallskip \noindent
{\bf Baselines.} The proposed MR\footnote{\url{https://gitee.com/mindspore/models/tree/master/official/recommend/mr}} and most existing debiasing methods are model-agnostic, which can be integrated into existing recommendation models for unbiased learning based on biased data. We follow prior work~\cite{CVIB,DIB} to adopt two of the most common
recommendation models as the backbones, i.e., matrix factorization
(MF)~\cite{MF} and neural collaborative filtering (NCF)~\cite{NCF}. Besides the naive estimator, we compare our method with the following state-of-the-art debiasing estimators: inverse propensity score (IPS) method~\cite{IPS}, the self-normalized IPS (SNIPS)~\cite{SNIPS}, doubly robust learning (DR)~\cite{DR}, doubly robust joint learning (DR-JL)~\cite{DRJL}, more robust doubly robust joint learning (MRDR-JL)~\cite{MRDR}, counterfactual variational information bottleneck (CVIB)~\cite{CVIB}, and debiased information bottleneck (DIB)~\cite{DIB}.

\smallskip \noindent
{\bf Evaluation Protocols and Experimental Details.}
We employ three widely used metrics to measure the prediction performance on the testing set for unbiased evaluation, including the mean square error (MSE), the area under the ROC curve (AUC), and the normalized discounted cumulative gain (nDCG). MSE measures the accuracy of the prediction, AUC measures the overall ranking performance on the testing set, and nDCG evaluates the ranking performance in a user-wise manner. All experiments are implemented on PyTorch~\cite{Pytorch} with Adam optimizer~\cite{Adam}, and grid search is used to choose the optimal set of hyper-parameters based on a validation set split from the training set. 

\subsection{Experiments on Real-world Datasets}
{\bf Datasets.} We consider two benchmark real-world datasets containing MNAR and MAR ratings, i.e., \textbf{Coat}\footnote{https://www.cs.cornell.edu/\textasciitilde schnabts/mnar/}~\cite{IPS} and \textbf{Yahoo\footnote{http://webscope.sandbox.yahoo.com/}}~\cite{Yahoo}, as existing work~\cite{IPS,DRJL}. Specifically, Coat has 6,960 five-star ratings from 290 Amazon Mechanical Turkers on an inventory of 300 coats in the training set, and 4,640 ratings collected from the 290 workers on 16 randomly selected coats in the testing set.
Yahoo includes a MNAR training set with 311,704 five-star ratings from 15,400 users and 1,000 songs, and a MAR testing set with 54,000 ratings from 5,400 users on 10 randomly selected songs. We follow prior works~\cite{IPS,DRJL,CVIB} to use the MNAR dataset for training and the MAR dataset for unbiased evaluation on both datasets. 

\begin{table*}[t]
    \centering
    \small
    \setlength{\tabcolsep}{3pt}
    \captionof{table}{Experimental results on Coat and Yahoo with MF and NCF as backbone models.}\label{tab:real}
\begin{threeparttable}  
\scalebox{0.98}{
\begin{tabular}{lcccc|cccc}
\toprule
Datasets   & \multicolumn{4}{c|}{Coat}           & \multicolumn{4}{c}{Yahoo}          \\
\cmidrule(r){1-1}  \cmidrule(lr){2-5} \cmidrule(lr){6-9}
Methods    & MSE    & AUC    & nDCG@5 & nDCG@10 & MSE    & AUC    & nDCG@5 & nDCG@10 \\
          \midrule
MF        & 0.2405 & 0.7028 & 0.6189 & 0.6858  & 0.2494 & 0.6806 & 0.6357 & 0.7640  \\
+IPS      & 0.2251 & 0.7152 & 0.6256 & 0.6934  & 0.2223 & 0.6831 & 0.6480 & 0.7665  \\
+SNIPS    & 0.2262 & 0.7082 & 0.6198 & 0.6861  & 0.1941 & 0.6834 & 0.6400 & 0.7648  \\
+DR       & 0.2325 & 0.7121 & 0.6246 & 0.6938  & 0.2106 & 0.6849 & 0.6580 & 0.7738  \\
+DR-JL    & 0.2312 & 0.7110 & 0.6209 & 0.6907  & 0.2175 & 0.6876 & 0.6458 & 0.7655  \\
+MRDR-JL  & 0.2301 & 0.7157 & 0.6325 & 0.6970  & 0.2169 & 0.6841 & 0.6465 & 0.7683  \\
+CVIB  & 0.2201 & 0.7247 & 0.6361 & 0.7030  & 0.2621 & 0.6856 & 0.6491 & 0.7718  \\
+DIB   & 0.2334 & 0.7104 & 0.6303 & 0.6986  & 0.2494 & 0.6832 & 0.6348 & 0.7633  \\
\midrule
+MR (Ours) & \textbf{0.2106} & \textbf{0.7356} & \textbf{0.6697} & \textbf{0.7343}  & \textbf{0.1920} & \textbf{0.6990} & \textbf{0.6709} & \textbf{0.7833}  \\
\midrule
\midrule
NCF        & 0.2116 & 0.7661 & 0.6293 & 0.7019  & 0.3318 & 0.6771 & 0.6532 & 0.7722  \\
+IPS      & 0.2002 & 0.7692 & 0.6362 & 0.7126  & 0.1706 & 0.6882 & 0.6630 & 0.7776  \\
+SNIPS    & \textbf{0.1920} & 0.7700 & 0.6313 & 0.7070  & 0.1697 & 0.6893 & 0.6687 & 0.7810  \\
+DR       & 0.2146 & 0.7523 & 0.6197 & 0.6908  & 0.1702 & 0.6890 & 0.6633 & 0.7779  \\
+DR-JL    & 0.2071 & 0.7612 & 0.6193 & 0.7021  & 0.2396 & 0.6811 & 0.6469 & 0.7653  \\
+MRDR-JL  & 0.2036 & 0.7629 & 0.6231 & 0.7011  & 0.2340 & 0.6834 & 0.6499 & 0.7681  \\
+CVIB  & 0.2060 & 0.7661 & 0.6244 & 0.6969  & 0.3055 & 0.6748 & 0.6701 & 0.7817  \\
+DIB   & 0.2030 & 0.7681 & 0.6300 & 0.7035  & 0.2849 & 0.7007 & 0.6757 & 0.7864  \\
\midrule
+MR (Ours) & 0.1945 & \textbf{0.7737} & \textbf{0.6393} & \textbf{0.7159}  & \textbf{0.1676} & \textbf{0.7026} & \textbf{0.7179} & \textbf{0.8112}  \\
\bottomrule
\multicolumn{9}{l}{$^\star$ The best results are highlighted in bold.}
\end{tabular}
}   
\end{threeparttable}
\vspace{-10pt}
\end{table*}

\begin{table*}[]
\centering
    \small
    \setlength{\tabcolsep}{3pt}
    \captionof{table}{Experimental results on the three semi-synthetic datasets, including ML-100k-1, ML-100k-2, and ML-100k-3, which have increasing level of exposure bias. We also present the relative performance drops of all methods on nDCG@10 in ML-100k-2 and ML-100k-3 compared with the corresponding results in ML-100k-1 to demonstrate the robustness of different models.}\label{tab:semi}
\begin{threeparttable}  
\scalebox{0.98}{
\begin{tabular}{lccc|ccc|ccc}
\toprule
Datasets & \multicolumn{3}{c|}{ML-100k-1} & \multicolumn{3}{c|}{ML-100k-2} & \multicolumn{3}{c}{ML-100k-3} \\
\cmidrule(r){1-1} \cmidrule(lr){2-4} \cmidrule(lr){5-7} \cmidrule(lr){8-10}
Methods  & MSE      & AUC     & nDCG@10  & MSE      & AUC     & nDCG@10  & MSE      & AUC     & nDCG@10  \\
         \midrule
Naïve    & 0.1188   & 0.6587  & 0.3223   & 0.1454   & 0.5700 & 0.2452 \tiny{(-23.9\%)}  & 0.2023  & 0.5105 & 0.2009 \tiny{(-37.7\%)}  \\
IPS      & 0.0913   & 0.7260  & 0.3517   & 0.1025   & 0.6321 & 0.2889 \tiny{(-17.9\%)}  & 0.2043  & 0.5144 & 0.2227 \tiny{(-36.7\%)}  \\
SNIPS    & 0.0914   & 0.7251  & 0.3526   & 0.1023   & 0.6318 & 0.2867 \tiny{(-18.7\%)}  & 0.1975  & 0.5126 & 0.2282  \tiny{(-35.3\%)} \\
DR    & 0.0997   & 0.6678  & 0.3204   & 0.1021   & 0.5926 & 0.2613 \tiny{(-18.4\%)}  & 0.1825  & 0.5186 & 0.2075 \tiny{(-35.2\%)}  \\
DR-JL    & \textbf{0.0896}   & 0.7296  & 0.3536   & 0.0975   & 0.6418 & 0.2875 \tiny{(-18.7\%)}  & 0.0991  & 0.5959 & 0.2838 \tiny{(-19.7\%)}  \\
MRDR-JL  & 0.0904   & 0.7268  & 0.3521   & 0.0980   & 0.6440 & 0.2919 \tiny{(-17.1\%)}  & 0.0987  & 0.6036 & 0.2823 \tiny{(-19.8\%)}  \\
CVIB  & 0.1036   & 0.6596  & 0.3151   & 0.1040   & 0.5538 & 0.2156 \tiny{(-31.6\%)}  & 0.2903  & 0.5050 & 0.1786 \tiny{(-43.3\%)}  \\
DIB  & 0.1190   & 0.6583  & 0.3243   & 0.1407   & 0.5756 & 0.2531 \tiny{(-22.0\%)}  & 0.1908  & 0.5117 & 0.1878 \tiny{(-42.1\%)}  \\
\midrule
MR (Ours) & 0.0902   & \textbf{0.7316}  & \textbf{0.3622}   & \textbf{0.0919} & \textbf{0.6968} & \textbf{0.3318} \tiny{(-8.4\%)}  & \textbf{0.0980}  & \textbf{0.6379} & \textbf{0.3100}  \tiny{(-14.4\%)} \\
\bottomrule
\multicolumn{10}{l}{$^\star$ The best results are highlighted in bold.}
\end{tabular}
}   
\end{threeparttable}
\vspace{-10pt}
\end{table*}

\smallskip \noindent
{\bf Performance.}
The experimental results are shown in Table~\ref{tab:real}. We have the following observations. 
Existing debiased estimators, including IPS, SNIPS, DR, DR-JL, MRDR-JL, CVIB and DIB, can effectively improve recommendation performance under biased training sets, since they all produce better results than the corresponding backbone models. DR methods only perform comparable to IPS based methods, which might be due to the inaccurate imputed errors. 
Impressively, our proposed MR estimator achieves significant improvements over the existing debiased methods in both Coat and Yahoo regardless of the backbone recommendation models. The reason can be attributed to its multiple robustness. By utilizing multiple different propensity models and imputation models, the MR estimator achieves state-of-the-art debiasing performance in terms of MSE, AUC and NDCG@K.

\subsection{Experiments on Semi-synthetic Datasets}
\textbf{Datasets.} We conduct experiments on semi-synthetic datasets constructed from \textbf{MovieLens 100K\footnote{https://grouplens.org/datasets/movielens/100k/} (ML-100k)} to evaluate the robustness of our proposed method. 
ML-100k is a dataset collected from a movie recommendation service with 100,000 MNAR ratings from 943 users and 1,682 movies. 
We generate three datasets based on it, denoted as ML-100k-1, ML-100k-2, and ML-100k-3, which have an increasing level of exposure bias. 

\smallskip \noindent
{\bf Performance.} The experimental results are presented in Table~\ref{tab:semi}. It can be found that our proposed MR estimator achieves the best results among all three semi-synthetic datasets, which demonstrates MR to be the most effective unbiased estimator compared with existing methods. Besides, MR shows more robust performance. Specifically, with the increase of exposure bias, the performance of IPS and SNIPS drops dramatically. 
For example, the AUC score of IPS drops up to 17.9\% and 36.7\% in ML-100k-2 and ML-100k-3, respectively, compared with that in ML-100k-1. 
Both DR-JL and MRDR-JL perform better than IPS and SNIPS due to their double robustness.
Our proposed MR has superior performance in all three datasets and shows significantly little performance drops with the increase of data bias, because of its multiple robustness for unbiased learning. 
\begin{table*}[]
\centering
    \small
    \setlength{\tabcolsep}{3pt}
    \captionof{table}{Performance of the MR method on Coat under different settings of imputation models, i.e., different numbers and types.}\label{tab:imput}
\begin{threeparttable}  
\scalebox{0.98}{
\begin{tabular}{lccc|lccc}
\toprule
Imputation Model  & MSE      & AUC     & nDCG@10  & Imputation Model  & MSE    & AUC     & nDCG@10 \\
\midrule
MF    & 0.2295   & 0.7209  & 0.7206   & MF   & 0.2295   & 0.7209  & 0.7206   \\
MF, MF   & 0.2252   & 0.7243  & 0.7301   & MF, MF   & 0.2252   & 0.7243  & 0.7301   \\
MF, MF, MF    & 0.2232   & 0.7332  & 0.7343   & NCF   & 0.2285   & 0.7230  & 0.7328   \\
MF, MF, MF, MF    & \textbf{0.2223}   & \textbf{0.7435}  & \textbf{0.7563}   & NCF, NCF   & \textbf{0.2093}   & \textbf{0.7381} & \textbf{0.7445}  \\
MF, MF, MF, MF, MF  & 0.2228   & 0.7421  & 0.7494   & MF, NCF   & 0.2143   & 0.7332 & 0.7325   \\
\bottomrule
\end{tabular}
}   
\end{threeparttable}
\end{table*}

\begin{table*}[t]
\centering
    \small
    \setlength{\tabcolsep}{3pt}
    \captionof{table}{
    Performance of the MR method under different numbers and types of propensity models on Coat dataset, where the imputation model and backbone prediction model both employ MF. 
    }\label{tab:indepth}
\begin{threeparttable}  
\scalebox{0.98}{
\begin{tabular}{lccc|lccc}
\toprule
Propensity Model  & MSE      & AUC     & nDCG@10  & Propensity Model  & MSE    & AUC     & nDCG@10 \\
\midrule
NB    & 0.2291   & 0.7219  & 0.7204   & NB   & \textbf{0.2291}   & \textbf{0.7219}  & 0.7204   \\
NB, NB-Uni   & 0.2269   & 0.7282  & 0.7322   & NB, NB   & 0.2293   & 0.7216  & 0.7195   \\
NB, NB-Uni, User    & \textbf{0.2228}   & \textbf{0.7370}  & \textbf{0.7347}   & NB, NB, NB   & 0.2293   & 0.7216  & \textbf{0.7206}   \\
\bottomrule
\end{tabular}
}   
\end{threeparttable}
\label{tab:ps}
\vskip -0.2in
\end{table*}

\subsection{In-depth Analysis of MR}
In this subsection, we conduct an in-depth analysis of the proposed MR estimator. 
\smallskip \noindent
{\bf Effect of Imputation Model.}
We study the effect of imputation model on the MR estimator by setting MR with different numbers and types of imputation models. Only one propensity model, i.e., the Naive Bayesian method, is used for all experiments. MF is used as the backbone prediction model, while both MF and NCF are used for imputations depending on specific configurations. 

As shown in the left side of Table~\ref{tab:imput}, when increasing the number of imputation models from one to five, all evaluation metrics first improve steadily and significantly, and then slightly drop. 
The best performance is achieved in the MR with four imputation models. It demonstrates that utilizing multiple imputation models instead of only one in the MR estimator can effectively improve model robustness and generalization.
Theoretically, as long as one of the imputation models estimate the prediction error accurately, the MR approach can achieve unbiased learning. 
However, it is worth noting that the MR method with five imputation models has slightly worse performance compared with that of fours. 
It can be interpreted as the additional MF model brings less improvement to the prediction accuracy than the noise. Therefore, the choice of the number of imputation models can be considered as a bias-noise trade-off, and the proper number of imputation models will lead to the optimal performance.

We further investigate the effect of the combination of different types of imputation models on the performance of MR methods, and the results are shown in the right side of Table~\ref{tab:imput}. First, MR methods with two MF and NCF imputations obtain an absolute increase in AUC scores of 3.4‰ and 5.1‰, respectively, compared to those methods with only one MF and NCF imputation model. Furthermore, although the mixture of MF and NCF performed competitively, the MR with two NCF imputations performs best. This suggests that deep recommendation models such as NCF perform better than simple models such as MF in terms of estimating prediction errors. Overall, the proper number of deep imputation models can help improve the debiasing performance of MR.

\begin{figure}[t]
\centering
\vspace{-2mm}
\includegraphics[width=0.4\textwidth]{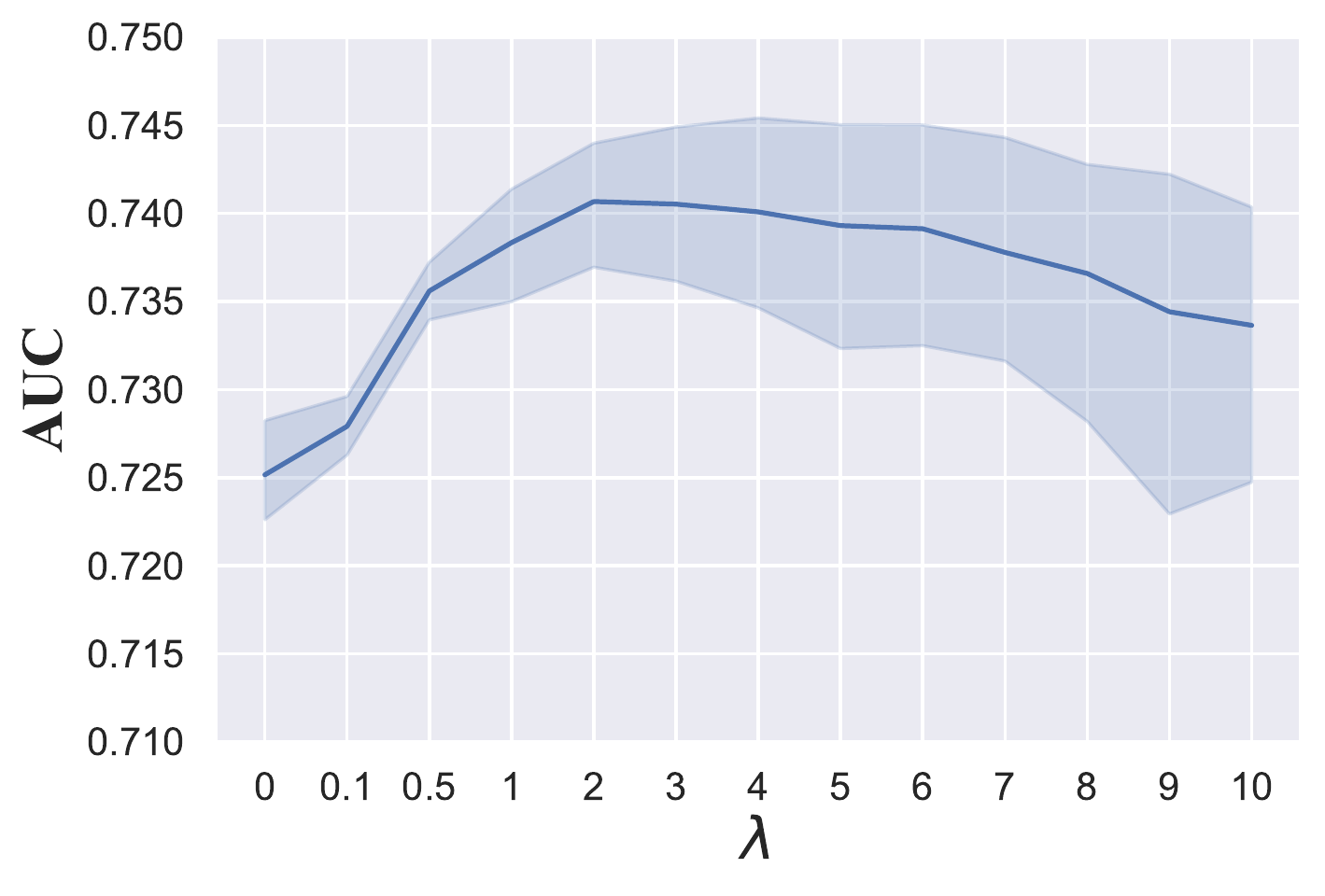} 
\vspace{-2mm}
\caption{AUC of MR with different settings of $\lambda$. The 90\% confidence of the results is shown with the shaded regions.}
\label{fig:lambda}
\vspace{-10pt}
\end{figure}

\smallskip \noindent
{\bf Effect of Propensity Model.} 
Table~\ref{tab:ps} shows the experimental results including different settings of propensity models on MR. 
Specifically, `NB', `NB-Uni', and `User' represent the Naive Bayes propensity estimator, uniform Naive Bayes propensity estimator, and user propensity estimator respectively. 
There are two experimental observations. Firstly, the MR method can achieve better performance by engaging several different types of propensity models as shown in the left part of the table. The reason is that different estimations of propensity scores can complement each other and contribute to the multiple robustness of MR beneficially. Secondly, employing the same type of several propensity models, such as two or three NB in the right part of table~\ref{tab:ps}, tends to harm the model performance. Since the additionally same type propensities provide no more useful information but bring redundant noises into the MR learning process.

\smallskip \noindent
{\bf Sensitivity of Hyper-parameter $\lambda$.}
Inspired by Theorem \ref{th5}, using a ridge regression form instead of the ordinary least square regression can further control the generalization error bound of MR. We further explore the effect of $\lambda$ in the equation (\ref{eq:eta}) on the debiasing performance of MR, where $\lambda$ is a hyperparameter of regularization strength. As shown in Figure~\ref{fig:lambda}, when varying $\lambda$ from 0 to 10.0, the AUC score increases from the beginning, and then declines a little,
 demonstrating that a proper $\lambda$ can further improve the performance of MR. Besides, it can effectively matigate the problem when computing $\boldsymbol{\hat \eta}$, especially $\sum_{  (u,i)\in \Tilde{\mathcal{O}}   } \boldsymbol{u}(x_{u,i}; {\alpha}, {\beta})\cdot  \boldsymbol{u}^{T}(x_{u,i}; {\alpha}, {\beta})$, runs into singular matrix.

\section{Conclusion}
In this paper, we propose a MR estimator to achieve unbiased learning in RS under weaker conditions.
Theoretical analysis demonstrates that the proposed estimator enjoys the property of multiple robustness, and shows that MR is actually an enhanced version of the DR estimator when the MR contains only one propensity model and one error imputation model. Further, we analyze the generalization error bound of MR and propose a novel multiple robust learning method with stabilization. We conduct extensive experiments on real-world and semi-synthetic datasets, which shows that the MR estimator can take advantage of multiple candidate propensity models and imputation models to significantly improve performance compared to the state-of-the-art debiasing approaches in RS. In future work, we will study the enhanced versions of MR estimator in RS and focus on sequential training for candidate models in MR.

\section{Acknowledgments}
This work was supported by the National Key R\&D Program of China (No. 2018YFB1701500 and No. 2018YFB1701503). We gratefully acknowledge the support from Mindspore\footnote{\url{https://www.mindspore.cn}}, CANN (Compute Architecture for Neural Networks) and Ascend AI Processor used for this research.

\bibliography{aaai23} 






\end{document}